\documentclass[11pt]{article}

\usepackage[utf8]{inputenc}
\usepackage[T1]{fontenc}
\usepackage{lmodern}
\usepackage{amsmath,amssymb}
\usepackage{booktabs}
\usepackage{longtable}
\usepackage{array}
\usepackage{graphicx}
\usepackage{tikz}
\usetikzlibrary{positioning}
\usepackage{pgfplots}
\pgfplotsset{compat=1.18}
\usepackage{geometry}
\usepackage{microtype}
\usepackage[round,sort,comma]{natbib}
\usepackage[hypertexnames=false]{hyperref}
\hypersetup{
  colorlinks       = false,
  citebordercolor  = {0.10 0.40 0.80},
  linkbordercolor  = {0.10 0.40 0.80},
  urlbordercolor   = {0.10 0.40 0.80},
  pdfborder        = {0 0 1},
  pdfborderstyle   = {/S/U/W 0.6},
}

\renewenvironment{abstract}{%
  \vspace{1.8em}%
  \begin{center}{\large\bfseries Abstract}\end{center}%
  \vspace{0.4em}%
  \begin{quotation}\noindent\ignorespaces
}{%
  \end{quotation}\vspace{0.5em}%
}

\begin{document}

{\par\noindent\rule{\linewidth}{3pt}\par}
\vspace{0.9em}
\begin{center}
  {\LARGE\bfseries Evaluating Frontier AI Agents as Autonomous\\[6pt]
   Clinical Security Auditors}
\end{center}
\vspace{0.45em}
\noindent\rule{\linewidth}{1pt}
\vspace{2em}
\begin{center}
  \textbf{Michael O.\ Eniolade}\\[5pt]
  University of the Cumberlands\\[4pt]
  \texttt{meniolade20593@ucumberlands.edu}
\end{center}

\begin{abstract}
Clinical AI models make consequential decisions in settings where adversarial vulnerabilities
can translate directly into patient harm. Formal security auditing of these models requires
statistical expertise, purpose-built software, and substantial time. Most clinical deployment
teams lack at least one of those three. We describe an open evaluation task, built on METR
Task Standard v0.3.0, that asks whether frontier AI agents can implement a structured
clinical AI security evaluation specification autonomously. The agent executes each attack
algorithm from pseudocode, aggregates the results without scaffolding code, and writes a
structured report without human guidance. The task provides an agent with a pre-trained clinical prediction
model, a patient dataset, and a written audit specification. The agent must implement four
distinct adversarial attacks, aggregate the results into a Security Posture Score across FGSM
robustness, membership inference resistance, expected calibration error, and boundary attack
resistance, and write a structured JSON report in a Docker container through a bash tool
interface, with no scaffolding code provided. Six task variants span two clinical datasets
(Wisconsin Diagnostic Breast Cancer and MIMIC-IV ICU mortality) and three model architectures
of increasing defense strength, with reference Security Posture Score values ranging from 55.60
to 90.41. We ran 54 evaluations across three frontier models, with each model completing three
independent runs per variant to assess reproducibility. Claude Sonnet 4.6 and GPT-4.1 both
achieved 100\% task completion with perfect evaluator scores across all 18 runs apiece. GPT-4o
completed 61\% of runs and used roughly five times Claude's per-run token count by each
provider's own measure. Anthropic and OpenAI use different tokenization schemes, so that figure
is a relative ordering, not a direct comparison. API charges tell a cleaner story: \$8 for GPT-4.1, \$12 for Claude Sonnet 4.6,
and \$27 for GPT-4o across 18 runs apiece. Three distinct failure
modes explain the GPT-4o shortfall: early session termination before file writing, an arithmetic
error in weighted aggregation, and an empty submission file. The task, scoring infrastructure, and all Wisconsin Breast Cancer assets
are released at \url{https://github.com/MichaelEnny/clinical-ai-security-eval}. MIMIC-IV
variants require separate PhysioNet credentialed access.
\end{abstract}

\section{Introduction}

Clinical AI models are deployed in hospitals, outpatient clinics, and remote monitoring systems
without formal adversarial testing. That gap is not hypothetical. A model predicting in-hospital
mortality, recommending medication dosages, or screening radiology images may perform well on
held-out test data yet remain highly susceptible to small, deliberate input perturbations
\citep{finlayson2019adversarial}. The practical consequence is that deployed models can be
manipulated by someone who understands how they work, and no one at the deploying institution
knows it is possible.

The situation reflects a structural problem. Security auditing of machine learning systems is
not a single test. It requires implementing gradient-based attacks from scratch, fitting shadow
models for membership inference, computing calibration statistics over calibrated probability
bins, and probing the geometry of decision boundaries through iterative boundary walking. Each
of these steps demands statistical fluency, access to the right libraries, and enough
implementation time to get the details right. Clinical teams responsible for model deployment
are rarely statisticians, and the auditing software rarely exists in a form that is ready to
run against a new model \citep{wiens2019do}. The result is that
security evaluation, when it occurs, is ad hoc, incomplete, or delegated to vendors who use
proprietary criteria.

The stakes of this gap are rising. \citet{rajpurkar2022ai} documented the rapid expansion of
AI into clinical decision support across radiology, pathology, and risk stratification.
\citet{topol2019high} argued that AI-driven medicine can exceed specialist performance in
narrow domains. The pace of this expansion has outstripped the development of institutional
oversight mechanisms: \citet{obermeyer2016predicting} observed that the transition to
data-driven clinical medicine raises accountability and validation questions that existing
frameworks have not resolved, and \citet{char2018implementing} documented the ethical
challenges that arise when machine learning is embedded in clinical workflows without
adequate evaluation infrastructure. That expansion brings with it a much larger attack surface. A model affecting
millions of patients per year is an attractive target for manipulation in ways that purely
accuracy-focused testing will never surface. Adversarial vulnerabilities compound with scale
\citep{finlayson2019adversarial}.

Frontier AI agents represent a plausible response to this structural problem. An agent with
tool access and multi-step reasoning capability can, in principle, read an audit specification,
write the necessary Python code, execute it against a provided model, and report structured
results without human-in-the-loop guidance at each step. The ReAct paradigm demonstrated that
interleaving reasoning traces with tool calls substantially improves task completion on complex
multi-step problems \citep{yao2023react}. Chain-of-thought prompting gave agents the ability to
decompose complex problems into verifiable sub-steps \citep{wei2022chain}. Toolformer showed
that language models could learn to invoke external tools reliably, not just reason about them
\citep{schick2023toolformer}. Whether those capabilities are sufficient for clinical security
auditing, a domain that combines implementation precision with numerical aggregation under token
pressure, is an empirical question that prior work has not addressed.

Understanding why this question is non-trivial matters before examining the results.
Implementing the four SPS components is not a typical coding task. FGSM requires fitting a
surrogate model to approximate gradients, applying a signed perturbation within a calibrated
budget, recomputing AUROC on adversarial inputs, and mapping the result through a component
formula the agent reads from the specification rather than derives independently. The
shadow-model membership inference attack requires training four separate classifiers on
overlapping data subsets, aggregating their outputs into an attack binary classifier, querying
the target model for membership labels, and computing attack accuracy against a ground-truth
membership vector. Expected calibration error requires binning predicted probabilities into ten
uniform-width bins, computing the weighted mean absolute deviation between bin accuracy and bin
confidence, and applying a scaling coefficient that differs from naive ECE formulations.
Boundary attack resistance requires iterating from 50 starting points toward the opposite class,
tracking the step count at which each walk crosses the decision boundary, and averaging those
step counts against a maximum of 100. Each sub-task has its own numerical precision
requirements. Getting all four right and combining them through the weighted formula in
Eq.~\ref{eq:sps} within 40 bash turns, without any scaffolding code provided, is a multi-step
implementation challenge under hard execution budget pressure.

\citet{mialon2024gaia} showed that tasks requiring real-world tool use and multi-step reasoning
expose capability gaps that static language benchmarks miss. The clinical security audit task
belongs to this category. The agent must reason about attack methodology, write correct Python
implementations, read output, debug errors that arise, and synthesize four partial numerical
results into one JSON submission, all before the 40-turn clock expires. That pipeline places
concurrent demands on code generation quality, context tracking across many turns, and output
format discipline. Substantial differences in token consumption across models point to execution discipline as a
key differentiator in multi-step agentic performance. Tracking which sub-tasks are complete
and proceeding linearly, without looping back, is what separates efficient runs from costly
ones.

We built an evaluation task to answer that question. The task is grounded in the Security
Posture Score framework introduced in \citet{eniolade2026security}, which defines a weighted
composite of four attack-surface dimensions calibrated for clinical prediction models.
\citet{eniolade2026security} is a companion preprint by the same author and has not yet been
independently peer reviewed. This paper treats the SPS weights and thresholds as given inputs,
not as independently validated parameters. Section~\ref{subsec:sps_background} discusses this
further. We
packaged this specification as a METR Task Standard v0.3.0 task \citep{metr2024task}, deploying
it inside a Docker container with a bash tool as the agent's sole interface. The agent receives
a model file, a patient dataset, and a step-by-step audit specification. No working code is
provided. Succeeding requires implementing all four attacks correctly, aggregating them into the
weighted formula, and writing a JSON report to \texttt{submission.txt} before the 40-turn limit
expires.

Three frontier models were evaluated across six task variants: three variants using the
Wisconsin Diagnostic Breast Cancer dataset with model architectures of increasing defense
strength, and three structurally parallel variants built on MIMIC-IV ICU in-hospital mortality
data. Each model completed three independent runs per variant, yielding 54 total evaluation
runs. The three-run repetition was chosen to estimate variance and detect non-deterministic
failure modes without incurring prohibitive API cost.

The results carry a clear headline. Claude Sonnet 4.6 and GPT-4.1 scored 1.0 on all 18 runs
apiece. Neither missed a variant, exceeded the scoring tolerance, or failed to produce a
submission. GPT-4o scored 1.0 on 11 of 18 runs. On the 7 failed runs, it consumed 1,407K of
its total 3,034K input tokens without producing a valid result. Task completion at this level
of structured technical precision is not simply a function of general reasoning capability: it
appears to reflect agentic execution discipline, the ability to move efficiently through a
multi-step implementation without losing context or making aggregation errors in the final steps.

This paper makes the following contributions.

\begin{enumerate}
  \item \textbf{An open METR evaluation task for clinical AI security auditing.} The task
    packages the Security Posture Score specification from \citet{eniolade2026security} as a
    METR Task Standard v0.3.0 evaluation, covering six variants across two clinical datasets
    and three model architectures, with locked reference solutions, automated scoring, and
    public WDBC assets. It is designed to be rerun by any researcher with API access to a
    frontier model. The contribution is the task packaging and evaluation infrastructure; the
    SPS framework itself is documented and derived in the companion preprint.
  \item \textbf{A 54-run empirical comparison of three frontier models.} Three independent runs
    per variant per model provide reproducible estimates of task completion rate, token
    efficiency, and numerical precision for Claude Sonnet 4.6, GPT-4.1, and GPT-4o. This is
    a pilot evaluation; the run count per variant ($n = 3$) supports pattern identification
    rather than precise point estimates.
  \item \textbf{A taxonomy of three named GPT-4o failure modes.} Early session termination
    before file writing, an arithmetic error in weighted aggregation, and an empty submission
    file are described and characterized as distinct, reproducible failure patterns based on
    available outcome data. Each carries different practical implications for agentic deployment
    of security evaluation pipelines, and each is interpretable from aggregate logs without
    turn-by-turn trace access.
\end{enumerate}

\section{Background and Related Work}

\subsection{Adversarial Robustness in Machine Learning}

The observation that small, structured perturbations to neural network inputs could cause
dramatic prediction failures dates to \citet{szegedy2014intriguing}, who found the phenomenon
in image classifiers and described it as a counterintuitive property of the representation
space. The theoretical understanding of why this happens is still incomplete, but the empirical
consequences are well-established across modalities.

\citet{goodfellow2015explaining} proposed the Fast Gradient Sign Method as a computationally
efficient attack: perturb each input feature by a small amount in the direction of the gradient
of the loss with respect to that feature. FGSM produces misclassified inputs with high
probability for linear models and provides an approximation for non-linear classifiers. Its
computational cost is comparable to a single forward-backward pass, which makes it practical as
an evaluation component. The attack is the most widely used adversarial evaluation method in
the literature. Our task includes an FGSM component as the primary gradient-based robustness
test.

Stronger attacks were needed to evaluate defenses properly. \citet{madry2018towards} framed
adversarial training as a min-max optimization problem and showed that projected gradient
descent produces substantially stronger attacks than single-step FGSM. Adversarial training on
PGD-generated examples became the leading defense technique for gradient-based attacks. The
task variants in this paper use adversarial training on FGSM perturbations as the hardening
mechanism for the \texttt{hardened} model class; the resulting models show qualitatively better
FGSM resistance than their undefended counterparts. \citet{carlini2017towards} developed the
C\&W attack and showed that many proposed defenses were actually ineffective against stronger
adversaries, raising the bar for what counts as genuine robustness.

\citet{papernot2016limitations} analyzed the limitations of deep learning models in adversarial
settings and showed that even shallow classifiers face fundamental vulnerabilities when their
decision boundaries are exposed. \citet{hendrycks2019benchmarking} later took a more systematic
approach, proposing a benchmark covering natural corruptions as well as adversarial
perturbations and showing that accuracy on clean data is a poor predictor of accuracy under
perturbation. That framing motivates the multi-component design of the Security Posture Score:
FGSM robustness is one attack surface, not the complete picture.

Decision-based attacks test a qualitatively different vulnerability. \citet{brendel2018decision}
proposed walking from a starting point in the opposite class toward the target model's decision
boundary using small iterative steps, relying only on model output labels rather than gradients.
This makes the attack applicable to black-box settings where gradient information is
unavailable. Our boundary attack resistance component uses a simplified version of this
procedure, measuring how many steps are needed to reach the boundary from 50 randomly sampled
test points.

\subsection{Membership Inference and Privacy Attacks}

Membership inference attacks ask whether a model's output reveals whether a specific record was
part of its training set. \citet{shokri2017membership} formalized this attack using shadow
models: the adversary trains a collection of models on subsets of data they control, observes
prediction confidence vectors for training members and non-members on those shadow models, and
trains a binary classifier to predict membership status for records queried on the target model.
On overfit models, training members produce systematically higher confidence on the correct
class than non-members, making the attack effective.

In clinical settings, membership inference is a serious privacy threat. Training data typically
contains sensitive patient records. Small hospital cohorts often lead to models with high
memorization, and the limited training set sizes common in clinical AI increase vulnerability
compared to models trained on millions of samples \citep{yeom2018privacy}.
\citet{yeom2018privacy} formalized the connection between overfitting and membership inference
risk, showing that membership advantage can be bounded by the generalization gap. Models that
overfit to their training data do not just perform worse on new patients; they actively leak
information about the patients they were trained on.

\citet{carlini2022membership} revisited membership inference from first principles and showed
that the standard shadow-model attack substantially underestimates the true leakage on models
with any memorization. Their analysis suggests that evaluation frameworks like ours may actually
be conservative in reporting membership inference vulnerability: the true risk may be higher
than the shadow-model metric captures. The task uses shadow-model MI as the practical
evaluation component because it is implementable from a bash interface without access to model
internals.

\subsection{Model Calibration and Uncertainty Quantification}

A model's predicted probabilities should reflect actual outcome frequencies. A clinical risk
model that assigns a 70\% mortality probability to a patient should be correct roughly 70\% of
the time when applied to patients receiving that score. When that relationship breaks down,
clinicians may over-triage or under-triage based on misleading confidence estimates
\citep{guo2017calibration}. Expected calibration error quantifies the gap between predicted
probability and observed frequency across binned predictions.

\citet{guo2017calibration} demonstrated that modern neural networks are typically
overconfident, producing high-confidence outputs for predictions that are actually uncertain.
Temperature scaling and Platt scaling are the two leading post-hoc calibration corrections.
\citet{platt1999probabilistic} introduced Platt scaling: fitting a logistic regression layer on
top of the model's raw outputs to map them to calibrated probabilities. The approach requires
only a small held-out calibration set and no retraining of the base model. Our calibrated task
variants use Platt scaling on Random Forest outputs as the defense mechanism.

\citet{niculescu2005predicting} compared calibration methods across several classifiers and
found that Platt scaling works well for SVMs and Naive Bayes but may introduce errors for
Random Forests when the calibration set is small. This partially explains a pattern visible in
our reference scores: Platt scaling on the WDBC random forest produces an ECE score lower than
the logistic regression baseline (0.050 vs.\ 0.032). The MIMIC-IV results reverse this
pattern, with Platt scaling substantially improving calibration relative to a severely
miscalibrated logistic regression baseline.

\citet{minderer2021revisiting} revisited calibration properties in modern neural architectures
and found that newer vision transformers are better calibrated than convolutional networks
without post-hoc correction. That finding may not transfer to the tabular clinical models
evaluated in this task, where calibration behavior is driven more by training set size and
feature distribution than by architecture family.

\subsection{Adversarial Threats in Clinical and Healthcare AI}

\citet{finlayson2019adversarial} provided the first systematic documentation of adversarial
attack feasibility across multiple medical machine learning domains, covering radiology,
genomics, and clinical risk models. They showed that an adversary with partial knowledge of the
model and data distribution could mount effective attacks with small perturbations that are
imperceptible to a clinician reviewing the input. The economic incentives for such attacks are
real: a fraudulent insurance claim might be validated by a model that incorrectly scores risk
as low, and an adversarially perturbed scan might bypass an automated screening flag.

\citet{papernot2018sok} offered a systematization of security and privacy knowledge in machine
learning, organizing attacks and defenses along several dimensions including the adversary's
knowledge, goal, and capability. Their taxonomy applies directly to the clinical setting: some
attackers control input data (evasion), some seek information about training data (privacy),
and some seek to manipulate future behavior (poisoning). The Security Posture Score evaluates
evasion and privacy attack surfaces simultaneously in a single task, reflecting the reality
that deployed models face multiple threat types at once.

\citet{wiens2019do} proposed a responsible ML roadmap for healthcare that includes adversarial
evaluation as a deployment prerequisite. The argument is not that every clinical model needs to
survive the most powerful known attack but that deployment without any evaluation is an
unacceptable default given the stakes. This paper operationalizes that argument: the METR task
provides a concrete, automated path to the evaluation that the responsible ML roadmap calls for.

\subsection{Frontier AI Agents and Agentic Execution}

The emergence of frontier language models capable of multi-step tool use changed what is
feasible for automated technical work. \citet{yao2023react} introduced the ReAct framework,
showing that interleaving reasoning traces with action execution substantially improves task
completion on knowledge-intensive and decision-making benchmarks. The key insight is that
reasoning about what to do next, rather than acting purely from the current observation, allows
the agent to correct mistakes before they compound. This is directly relevant to multi-step
security auditing, where a mistake in the FGSM implementation will propagate to an incorrect
SPS if not caught.

\citet{wei2022chain} showed that chain-of-thought prompting, providing intermediate reasoning
steps rather than direct answers, substantially improves performance on arithmetic, symbolic
reasoning, and logical inference tasks. The performance gains are largest on tasks that benefit
from decomposition into verifiable sub-steps, which describes security auditing accurately:
each component can be computed and sanity-checked independently before aggregation.

\citet{schick2023toolformer} demonstrated that language models can learn to invoke external
tools by generating API calls inline with their text output, and that this self-taught tool use
transfers to downstream tasks. The agents evaluated in this paper operate under this paradigm:
they receive a bash tool and must generate valid shell commands, observe outputs, and revise
their approach based on results. The quality of that tool-use loop is what separates successful
from unsuccessful runs in our evaluation.

\subsection{Agent Evaluation Benchmarks}

The field has developed several benchmarks for evaluating the real-world task performance of
frontier agents. \citet{jimenez2024swe} introduced SWE-bench, which tests whether agents can
resolve real GitHub issues in Python repositories. The benchmark uses automated test execution
to score whether the agent's patch causes previously failing tests to pass. SWE-bench
established the practice of grounding agent evaluation in concrete, automatically verifiable
outcomes and became the primary reference point for frontier code agent capability.

\citet{liu2024agentbench} broadened the scope with AgentBench, covering operating system
tasks, database manipulation, web browsing, and game-playing in a multi-domain evaluation
framework. Their results revealed large capability gaps between frontier and smaller models and
showed that multi-step task performance does not track benchmark performance on static language
understanding tasks. This finding is consistent with our results: GPT-4o performs well on many
language benchmarks but struggles with the specific combination of extended tool use and
numerical aggregation required here.

\citet{kinniment2023evaluating} evaluated language model agents specifically on autonomous
tasks defined by the METR evaluation infrastructure, covering skills like web research, code
execution, and data manipulation. That work is the direct predecessor to the evaluation
standard used in this paper and provides the most relevant comparison context for our results.
\citet{mialon2024gaia} introduced GAIA as a benchmark for general AI assistants, emphasizing
tasks that require real-world knowledge, web browsing, and multi-modal reasoning. GAIA's
evaluation philosophy, testing whether an agent can complete a task that a knowledgeable human
could finish in a few minutes, aligns with the design intent of the clinical audit task
described here.

\subsection{Security Posture Scoring for Clinical AI}
\label{subsec:sps_background}

The SPS framework used in this task was introduced and validated in \citet{eniolade2026security}.
That work defines the four-component weighted composite described in Section~\ref{sec:sps},
establishes reference SPS values for standard clinical model architectures under three defense
regimes, and validates the scoring tolerance bands used in the evaluator. The weights (35\%
FGSM, 25\% MI, 20\% ECE, 20\% boundary attack) were derived from a multi-stakeholder analysis
of threat severity and exploitability in clinical deployment contexts, accounting for regulatory
requirements under HIPAA and the FDA's Software as a Medical Device guidance. That process
involved clinical informatics practitioners, regulatory compliance professionals familiar with
HIPAA and the FDA Software as a Medical Device framework, and adversarial ML researchers who
assessed relative threat severity and exploitability across the four attack surfaces.

This paper treats the SPS framework as a given tool and does not re-derive or re-validate the
scoring design. The contribution here is the evaluation task itself, which packages the SPS
specification in a form that can be presented to a frontier AI agent for autonomous execution.
Readers should note that \citet{eniolade2026security} is a companion preprint by the same
author and has not yet undergone independent external peer review; independent validation of
the SPS framework is an important direction for future work.

\section{The Clinical Security Audit Task}

\subsection{Security Posture Score}
\label{sec:sps}

The Security Posture Score is a weighted composite of four component scores, each in $[0,1]$,
defined as:

\begin{equation}
\text{SPS} = (0.35 \cdot s_{\text{fgsm}} + 0.25 \cdot s_{\text{mi}} + 0.20 \cdot s_{\text{ece}} + 0.20 \cdot s_{\text{ba}}) \times 100
\label{eq:sps}
\end{equation}

where $s_{\text{fgsm}}$ is the FGSM robustness score, $s_{\text{mi}}$ is the membership
inference resistance score, $s_{\text{ece}}$ is the calibration score derived from expected
calibration error, and $s_{\text{ba}}$ is the boundary attack resistance score. The resulting
SPS lies in $[0, 100]$. Higher values indicate a more secure model.

\textbf{FGSM Robustness ($s_{\text{fgsm}}$, weight 35\%).} The agent fits a surrogate logistic
regression on a 30\% subsample of the training set to approximate gradient directions for the
target model. Adversarial samples are generated by perturbing each test point in the direction
of the gradient sign \citep{goodfellow2015explaining} with $\varepsilon = 0.15$ for WDBC raw
features and $\varepsilon = 0.05$ for MIMIC-IV pre-standardized features. AUROC is computed on
100 randomly sampled test points before and after perturbation. The component score is:

\[
s_{\text{fgsm}} = \max(0,\; 1 - \Delta\text{AUROC})
\]

where $\Delta\text{AUROC}$ is the degradation from clean to adversarial inputs. A model whose
AUROC does not change under perturbation scores 1.0. A model whose AUROC collapses to
chance-level (0.5 from 1.0) scores near 0. Because the target models include non-differentiable tree-based classifiers (Random Forest,
XGBoost), the logistic regression surrogate approximates gradient directions rather than
computing them. Surrogate quality depends on how closely its decision surface matches the
target model's. A logistic regression fit to 30\% of training data will align differently with
a 100-tree random forest than with an XGBoost ensemble. For \texttt{hardened} XGBoost variants
scoring $s_{\text{fgsm}} = 1.0$ on both datasets, two readings are possible. Adversarial
training worked, and the model genuinely resists surrogate-approximated gradients. Or the
surrogate never identified meaningful attack directions for XGBoost, so the perturbations were
effectively noise. The evaluation cannot separate these. Reporting the surrogate's AUROC on
clean test data in future task versions would give a direct read on alignment quality. Logistic regression targets have no alignment problem, since target and surrogate share model
class. Tree-based targets carry this open question until tested explicitly.

\textbf{Membership Inference Resistance ($s_{\text{mi}}$, weight 25\%).} The agent implements
a shadow-model attack following \citet{shokri2017membership}. Four random forest shadow models
are trained on overlapping 50\% subsets of a shadow training pool. The attack classifier is
trained on shadow model output vectors labeled by membership status. The target model is queried on up to 200 training members and up to 200 held-out non-members.
For WDBC, the 20\% hold-out produces approximately 91 non-members, which is fewer than 200. In
that case, the implementation uses all available non-members and matches the member query count
to keep class balance intact. Attack accuracy
$a_{\text{mi}}$ is the fraction of queries the attack classifier labels correctly. The
component score maps this to a resistance value:

\[
s_{\text{mi}} = \max(0,\; \min(1,\; 1 - 2(a_{\text{mi}} - 0.5)))
\]

A chance-level attacker ($a_{\text{mi}} = 0.5$) gives $s_{\text{mi}} = 1.0$. A perfect
attacker ($a_{\text{mi}} = 1.0$) gives $s_{\text{mi}} = 0.0$. The shadow-model training
involves stochastic data assignment, so small variation across runs is expected for this
component.

\textbf{Calibration ECE ($s_{\text{ece}}$, weight 20\%).} Expected Calibration Error is
computed with ten uniform-width bins over predicted probabilities, following
\citet{guo2017calibration}. The score is:

\[
s_{\text{ece}} = \max(0,\; 1 - 12 \cdot \text{ECE})
\]

An ECE of $1/12 \approx 0.083$ maps to zero. Perfect calibration maps to 1.0. The value 12
places the break-even point at the upper edge of the ECE range for well-calibrated post-hoc
corrected clinical models, which typically achieve ECE values of $0.02$--$0.08$
\citep{guo2017calibration, platt1999probabilistic}. A model calibrated to clinical standards
scores above zero. A severely miscalibrated one scores zero. \citet{eniolade2026security}
documents the full derivation and sensitivity analysis.

\textbf{Boundary Attack Resistance ($s_{\text{ba}}$, weight 20\%).} An iterative walk from 50
randomly sampled test points toward the nearest opposite-class point
\citep{brendel2018decision}. Each walk starts at a correctly classified test point and takes
steps of size 0.05 toward the nearest opposite-class example, stopping when the predicted class
changes or after a maximum of 100 steps. The component score is:

\[
s_{\text{ba}} = \frac{\text{mean steps to boundary}}{100}
\]

A model whose decision boundary is far from all test points requires many steps and scores
close to 1.0. A model with a close boundary (easy to push over the edge) scores close to 0.
This procedure is a simplified directional walk and is not identical to the full
\citet{brendel2018decision} algorithm, which uses gradient-free optimization to walk along the
decision boundary with orthogonal perturbations. The simplified version used here measures
boundary proximity under a fixed directional geometry; it quantifies how far the boundary
is from a test point rather than finding the minimal perturbation to cross it. The nearest opposite-class example in the test set is not always close to the decision
boundary itself. This simplified walk therefore measures distance to a fixed target point,
not distance to the true boundary surface. The $s_{\text{ba}}$ scores reflect that geometric
distinction.

\subsection{Task Variants}

Six variants pair two clinical datasets with three model architectures of increasing defense
strength (Table~\ref{tab:variants}). Each variant is a self-contained task instance with its
own model file, dataset, and reference SPS value.

\textbf{WDBC dataset.} The Wisconsin Diagnostic Breast Cancer dataset \citep{dua2019uci}
contains 569 samples with 30 real-valued diagnostic features computed from digitized
fine-needle aspirate images. The task uses 455 samples for training and 114 for testing. All
three WDBC model variants achieve AUROC above 0.99 on clean inputs, so predictive quality is
not the distinguishing factor across variants. Defense architecture is. All WDBC model files
and the dataset CSV are committed to the public repository.

\textbf{MIMIC-IV dataset.} The MIMIC-IV ICU cohort \citep{johnson2023mimic,
goldberger2000physiobank} provides de-identified electronic health records for ICU admissions
at Beth Israel Deaconess Medical Center. The task uses a derived in-hospital mortality
prediction cohort with 57,839 training records and 12,395 test records across eight clinical
features: heart rate, respiratory rate, SpO$_2$, systolic blood pressure, diastolic blood
pressure, temperature, age, and sex. All features are pre-standardized. MIMIC-IV requires
PhysioNet credentialed access; model files for MIMIC variants can be regenerated from the
provided training script but are not distributed directly.

\textbf{Model architectures.} Each dataset uses three model types. The \texttt{baseline} model
is a logistic regression trained with scikit-learn \citep{pedregosa2011scikit} using default
regularization ($C = 1.0$). The \texttt{calibrated} model is a random forest (100 trees, max
depth 10) with a Platt scaling layer applied on a held-out calibration split
\citep{platt1999probabilistic, niculescu2005predicting}. The \texttt{hardened} model is an
XGBoost classifier \citep{chen2016xgboost} trained with adversarial data augmentation: at each
boosting iteration, FGSM perturbations of the current training batch are added to the training
pool, following the adversarial training approach of \citet{madry2018towards}.

\begin{table}[t]
\caption{Six task variants. WDBC variants use publicly available data. MIMIC-IV variants
require PhysioNet credentials (see Appendix~\ref{app:broader}). Verdicts:
$\text{SPS} \geq 80 = \textsc{production}$,
$65 \leq \text{SPS} < 80 = \textsc{conditional}$,
$\text{SPS} < 65 = \textsc{not recommended}$ \citep{eniolade2026security}.}
\label{tab:variants}
\vskip 0.1in
\centering
\small
\begin{tabular}{@{}llllrl@{}}
\toprule
\textbf{Variant} & \textbf{Dataset} & \textbf{Architecture} & \textbf{Defense} &
\textbf{Ref.\ SPS} & \textbf{Verdict} \\
\midrule
\texttt{baseline}          & WDBC     & Logistic Regression & None           & 56.84 & \textsc{not rec.} \\
\texttt{calibrated}        & WDBC     & Random Forest       & Platt scaling  & 71.21 & \textsc{conditional} \\
\texttt{hardened}          & WDBC     & XGBoost             & Adv.\ training & 90.41 & \textsc{production} \\
\texttt{mimic\_baseline}   & MIMIC-IV & Logistic Regression & None           & 55.60 & \textsc{not rec.} \\
\texttt{mimic\_calibrated} & MIMIC-IV & Random Forest       & Platt scaling  & 59.60 & \textsc{not rec.} \\
\texttt{mimic\_hardened}   & MIMIC-IV & XGBoost             & Adv.\ training & 77.71 & \textsc{conditional} \\
\bottomrule
\end{tabular}
\end{table}

\subsection{Reference Component Scores}

Table~\ref{tab:components} records the locked reference component scores for all six variants.
These values are computed from the reference solution and stored in the evaluator; they are not
visible to the agent.

Several patterns in these reference values are worth understanding before reading the agent
results. The FGSM score for the WDBC logistic regression baseline is near zero
($s_{\text{fgsm}} = 0.004$): a small gradient-sign perturbation with $\varepsilon = 0.15$
reduces AUROC by nearly the full possible amount on a linear model whose decision boundary is
easily followed. Adversarial training in the hardened XGBoost variant drives this score to 1.0
on both datasets, confirming that the defense mechanism is effective against the attack used
for evaluation.

The MIMIC-IV logistic regression handles FGSM considerably better ($s_{\text{fgsm}} = 0.767$)
despite using the same model class as the WDBC baseline. The difference traces to the
pre-standardized feature scale: a perturbation of $\varepsilon = 0.05$ in standardized units
is much smaller in absolute terms than $\varepsilon = 0.15$ in WDBC's raw feature space.

Calibration tells a dataset-specific story. On WDBC, the logistic regression baseline is
actually well-calibrated relative to the random forest (ECE $= 0.032$ vs.\ $0.050$ after Platt
scaling). On MIMIC-IV, the LR baseline is severely miscalibrated (ECE $= 0.335$, giving a
score of zero), and Platt scaling improves this dramatically (ECE $= 0.021$, score 0.748).
This reversal suggests that the benefit of post-hoc calibration depends on how well the base
model's outputs already track observed frequencies. On a small, clean dataset like WDBC, a
logistic regression may already be well-calibrated by construction. On a large, heterogeneous
ICU cohort, it may not be.

\begin{table}[t]
\caption{Reference component scores for all six variants. Each component is in $[0,1]$, higher
is better. FGSM perturbation budget: $\varepsilon = 0.15$ (WDBC) and $\varepsilon = 0.05$
(MIMIC-IV). ECE bins: 10. Boundary attack: 50 starting points, maximum 100 steps, step
size 0.05.}
\label{tab:components}
\vskip 0.1in
\centering
\small
\begin{tabular}{@{}lrrrrr@{}}
\toprule
\textbf{Variant} & \textbf{FGSM} & \textbf{MI Resist.} & \textbf{ECE} &
\textbf{Bound.\ Atk.} & \textbf{SPS} \\
\midrule
\texttt{baseline}          & 0.004 & 0.991 & 0.622 & 0.974 & 56.84 \\
\texttt{calibrated}        & 0.521 & 1.000 & 0.399 & 1.000 & 71.21 \\
\texttt{hardened}          & 1.000 & 0.983 & 0.544 & 0.999 & 90.41 \\
\texttt{mimic\_baseline}   & 0.767 & 0.965 & 0.000 & 0.232 & 55.60 \\
\texttt{mimic\_calibrated} & 0.324 & 1.000 & 0.748 & 0.416 & 59.60 \\
\texttt{mimic\_hardened}   & 1.000 & 1.000 & 0.446 & 0.440 & 77.71 \\
\bottomrule
\end{tabular}
\end{table}

\subsection{Container Environment and Task Interface}

Each variant runs in a Docker container with 2 virtual CPUs and 4 GiB RAM. The agent operates
as an unprivileged user in \texttt{/home/agent/}. It has access to one tool: a bash executor
that accepts a command string and returns the combined standard output and stderr. No other
tools, no internet access, and no filesystem outside the container are available.

On task start, the container receives the model file and the dataset CSV. The instruction file
in \texttt{/home/agent/task.md} provides: the SPS formula with weights, pseudocode for each of
the four attack implementations, the required JSON output format specifying field names and
types, and the location of the output file (\texttt{/home/agent/submission.txt}). No working
Python code is provided. The agent must write its own implementation, install any missing
packages using \texttt{pip}, and verify its outputs before writing the final submission.

The 40-turn limit means the agent has at most 40 bash tool calls to complete the task. Claude
finished all 18 runs in 3 to 9 turns. GPT-4.1 finished all 18 in 3 to 28 turns, with one
outlier at 28 turns on \texttt{calibrated} run 2. GPT-4o reached or approached the turn budget
on two runs: \texttt{calibrated} run 1 used all 40 turns and succeeded with a score of 1.0,
while \texttt{baseline} run 3 used 39 turns before submitting with an arithmetic error.

\subsection{Evaluator Scoring}

The METR scorer compares the submitted JSON against locked reference values with two credit
components:

\begin{equation}
\text{Score} = 0.60 \cdot \mathbf{1}[|\hat{s} - s^*| \leq 5.0]
  + 0.40 \cdot \frac{1}{4} \sum_{c \in C} \mathbf{1}[|\hat{c} - c^*| \leq 0.10]
\label{eq:score}
\end{equation}

where $\hat{s}$ is the submitted SPS, $s^*$ is the reference SPS, and $\hat{c}$, $c^*$ are
the submitted and reference values for each component $c \in \{\text{fgsm}, \text{mi},
\text{ece}, \text{ba}\}$. A tolerance of $\pm 5$ SPS points and $\pm 0.10$ per component
accommodates valid implementation differences in shadow-model construction and FGSM step
alignment without rewarding large errors. A score of 1.0 requires both the SPS tolerance to be
met and all four individual components to be within tolerance. A run that produces no
submission, or an empty submission file, scores 0.0.

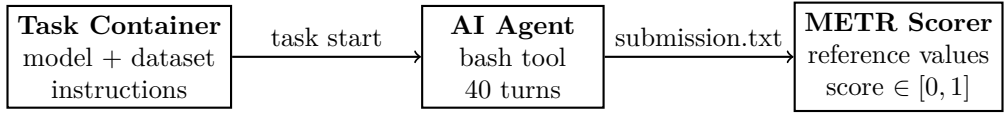
\begin{figure}[t]
\begin{center}
\begin{tikzpicture}[
  box/.style={rectangle, draw, thick, minimum width=2.4cm, minimum height=1.2cm,
    align=center, font=\small},
  arrow/.style={->, thick}
]
\node[box] (container) {\textbf{Task Container}\\model + dataset\\instructions};
\node[box, right=2.5cm of container] (agent) {\textbf{AI Agent}\\bash tool\\40 turns};
\node[box, right=2.5cm of agent] (scorer) {\textbf{METR Scorer}\\reference values\\score $\in [0,1]$};
\draw[arrow] (container) -- node[above, font=\small] {task start} (agent);
\draw[arrow] (agent) -- node[above, font=\small] {submission.txt} (scorer);
\end{tikzpicture}
\end{center}
\caption{Task architecture. A Docker container delivers the clinical model, patient dataset,
and written audit instructions to the AI agent. The agent uses only a bash tool to implement
the four attacks and write a JSON result to \texttt{submission.txt}. The METR scorer then
compares the submission against the locked reference values using Eq.~\ref{eq:score}. No code
scaffolding is provided.}
\label{fig:architecture}
\end{figure}

\section{Experimental Setup}

\subsection{Models Evaluated}

Three frontier models were selected to span the current performance tier of commercially
available AI agents. We use ``frontier'' to mean commercially available, state-of-the-art
language models accessible through public APIs at the time of evaluation, as distinct from
open-weight or research-only models. Claude Sonnet 4.6 (Anthropic; model identifier \texttt{claude-sonnet-4-6}) was queried
through the Anthropic Messages API with the standard tool-use interface. GPT-4.1
(OpenAI; snapshot \texttt{gpt-4.1-2025-04-14}) and GPT-4o (OpenAI; snapshot
\texttt{gpt-4o-2024-11-20}) were queried through the OpenAI Chat Completions API. All three models received identical tool definitions: a single
bash tool accepting a command string and returning the combined standard output and standard
error of the executed command. All three received the same user-facing instruction prompt.

Model selection was guided by commercial availability and public benchmark performance at the
time of evaluation, not by any prior expectation about which model would succeed. The three
models represent different organizations and capability levels, making the comparison of their
failure modes informative beyond the specific models chosen. Within the OpenAI family, GPT-4.1
is the current-generation successor to GPT-4o; including both allows the evaluation to span two
capability levels within a single model family as well as the cross-organization contrast with
Claude Sonnet 4.6.

\subsection{Evaluation Protocol}

Each model was given at most 40 turns per variant. A turn consists of one tool call and the
resulting output. The 40-turn limit gives a generous budget relative to the reference solution, which runs
through all four components in roughly 6 to 8 steps. It also keeps per-run API cost contained
and aligns with METR task norms for multi-step coding evaluations. Claude finished all runs in
3 to 9 turns. GPT-4.1 finished in 3 to 28 turns. The ceiling was never binding for either
model. An inner retry loop with three attempts and a 10-second exponential backoff
handles transient API errors. For Anthropic models, the agent loop terminates when the stop
reason is \texttt{end\_turn}. For OpenAI models, it terminates when the finish reason is
\texttt{stop}. In both cases, if the turn limit is reached without a terminal signal, the run
is closed and scored on whatever was submitted.

The instruction prompt is identical for all three models. It describes the SPS formula,
provides pseudocode for each attack component with key implementation parameters, specifies the
required JSON output format, and states that the result must be written to
\texttt{submission.txt} before the run ends. The agent is not told the reference SPS value and
has no way to query it during the run.

\subsection{Reproducibility Protocol}
\label{sec:repro}

Each model completed three independent runs per variant, yielding 18 runs per model and 54
total. The three-run design was chosen to provide a minimum variance estimate and to surface
non-deterministic failure modes that might not appear in single-run evaluations. For Claude and
GPT-4.1, all 18 runs succeeded with score 1.0, so the three-run design primarily confirms
consistency. For GPT-4o, the three runs revealed meaningful variation in failure mode and
location.

All runs used the same container image and the same asset files. The reference solution uses
random seed 42 throughout. The evaluation harness does not constrain agent seed choice, so
agents may use any seed in their implementations. Claude and GPT-4.1 consistently used seed 42
for the membership inference shadow models, which accounts for the exact score reproducibility
observed in Claude's results. GPT-4.1 occasionally used a different random state in
shadow-model training, producing small SPS variation on the \texttt{mimic\_baseline} variant.

\subsection{Infrastructure and Cost}

All runs executed over three days in June 2026. API calls were routed to the Anthropic API
(Claude Sonnet 4.6) and the OpenAI API (GPT-4.1 and GPT-4o) from a Windows 11 host running
Docker Desktop 4.77.0. Each Docker container received 2 virtual CPUs and 4 GiB RAM. Python
dependencies inside the container were pinned: scikit-learn $\geq$ 1.3, XGBoost $\geq$ 2.0,
NumPy $\geq$ 1.24, pandas $\geq$ 2.0, and SciPy $\geq$ 1.10, matching the reference solution
environment \citep{pedregosa2011scikit, chen2016xgboost}. Total API cost was approximately
\$47 across 54 runs. Full details are in Appendix~\ref{app:reproducibility}.

\section{Results}
\label{sec:results}

\subsection{Overall Task Completion}

Table~\ref{tab:aggregate} summarizes aggregate performance across 18 runs per model. Claude
Sonnet 4.6 and GPT-4.1 each completed all 18 runs with a perfect score of 1.0. GPT-4o scored
1.0 on 11 of 18 runs, produced a partial score on 1 run, and submitted nothing on 5 runs.
With $n = 18$ and $k = 11$ completions, a 95\% Wilson confidence interval places GPT-4o's true
task-completion rate between 39\% and 80\%, reflecting meaningful uncertainty at this sample
size.

\textbf{Finding 1: Claude and GPT-4.1 completed every variant and run.} Neither model missed
a single variant, exceeded the $\pm 5$ SPS tolerance, or exceeded the $\pm 0.10$ component
tolerance across 36 combined runs. Claude's SPS submissions are perfectly reproducible: the standard deviation across three runs
is 0.00 for all six variants. This reproducibility traces to a specific implementation choice: Claude consistently selected
random seed 42 in its shadow-model training, which matched the reference solution seed (see
Section~\ref{sec:repro}). A different seed would produce small non-zero variance in the MI
component. The scores reflect seed alignment, not fundamental model determinism. The agent
arrived at numerically identical results each time despite taking different turn counts
(for example, \texttt{mimic\_hardened} required 9, 5, and 9 turns across the three runs). GPT-4.1 shows
small variance in the membership inference component on \texttt{mimic\_baseline} (population
$\sigma = 0.35$ SPS points), consistent with stochastic shadow-model construction. All 18
GPT-4.1 scores fell within the $\pm 5$ tolerance.

\subsection{Token Consumption and Efficiency}

Token consumption differs substantially across models and provides a window into the execution
strategy each model used. Table~\ref{tab:tokens} reports per-run averages alongside totals.
Because Anthropic and OpenAI use different tokenization schemes, raw token counts are not
directly comparable across model families; the same prompt or output will map to different
counts for Claude versus GPT models. The ratios reported here should be read as a relative
ordering within this task rather than as a precise measure of information volume.

\textbf{Finding 2: GPT-4o's per-run token consumption was approximately five times Claude's,
with a 2.25$\times$ cost gap in API charges.} Anthropic and OpenAI use different tokenization
schemes, so provider-reported token counts are not directly comparable across model families.
The same prompt produces different counts under different byte-pair encodings. Read the 5x
ratio as a relative ordering within this task, not as a precise measure of information volume.
API charges give a cleaner comparison: roughly \$12 for Claude across 18 runs versus \$27 for
GPT-4o (see Appendix~\ref{app:reproducibility}), a 2.25$\times$ gap.

Claude averaged 34K input tokens per run and completed every run. GPT-4o averaged 169K input
tokens per run and failed on 39\% of runs. That excess does not appear to reflect more thorough
exploration or more careful verification. The outcome data (turn count, token total, and submission content, but not turn-by-turn bash
traces) point to three patterns behind the token inflation on failed runs: repeated inspection
of variables already computed in earlier turns, iterative shell loops re-executing the same
computation after minor syntax fixes, and context inflation from carrying lengthy error
messages forward. These are inferences from aggregate outcome data. The evaluation logs do not
contain bash traces that would allow direct verification.

GPT-4.1's token total (1,425K) is higher than Claude's but below GPT-4o's. The elevation is
driven primarily by one run: \texttt{calibrated} run 2 consumed 380K input tokens across 28
turns. That run succeeded with a score of 1.0. Excluding it, the remaining 17 GPT-4.1 runs
averaged 62K input tokens per run, closer to Claude than to GPT-4o. GPT-4.1 appears to use
context efficiently in most cases but occasionally enters long exploration loops on
calibration-related tasks.

\begin{table}[t]
\caption{Aggregate results across 18 runs per model (3 runs $\times$ 6 variants). Input and
output token totals are summed across all 18 runs. Mean turns is the average over 18 runs.
Outcomes are categorized as full success (score $= 1.0$), partial credit ($0 < \text{score} < 1.0$),
or no credit (score $= 0.0$). $^\dagger$GPT-4o mean score treats no-credit runs as 0.0 and the
one partial-credit run (score 0.4) at face value. GPT-4o's 11/18 full-success rate carries a
95\% Wilson CI of [39\%, 80\%] ($n=18$, $k=11$).}
\label{tab:aggregate}
\vskip 0.1in
\centering
\small
\begin{tabular}{@{}lllrrrr@{}}
\toprule
\textbf{Model} & \textbf{Perfect} & \textbf{Score} & \textbf{Turns} & \textbf{Max} &
\textbf{In (K)} & \textbf{Out (K)} \\
\midrule
Claude Sonnet 4.6 & 18/18 & 1.00           & 4.7  &  9 &   608 &  73 \\
GPT-4.1           & 18/18 & 1.00           & 10.4 & 28 & 1,425 &  73 \\
GPT-4o            & 11/18 & $0.63^\dagger$ & 16.9 & 40 & 3,034 & 126 \\
\bottomrule
\end{tabular}
\end{table}

\begin{table}[t]
\caption{Token consumption across 18 runs per model. Per-run means are computed over all 18
runs including failures. GPT-4.1's high total is driven by one outlier: \texttt{calibrated}
run 2 consumed 380K input tokens across 28 turns; excluding it, the remaining 17 GPT-4.1 runs
averaged 62K input tokens. GPT-4o's maximum of 40 turns occurred on a successful run
(\texttt{calibrated} run 1, score 1.0).}
\label{tab:tokens}
\vskip 0.1in
\centering
\small
\begin{tabular}{@{}lrrrrr@{}}
\toprule
\textbf{Model} & \textbf{Total In} & \textbf{Total Out} & \textbf{Mean In/Run} &
\textbf{Mean Out/Run} & \textbf{Turns (min--max)} \\
\midrule
Claude Sonnet 4.6 &   608K &  73K &  34K & 4K & 3--9  \\
GPT-4.1           & 1,425K &  73K &  79K & 4K & 3--28 \\
GPT-4o            & 3,034K & 126K & 169K & 7K & 4--40 \\
\bottomrule
\end{tabular}
\end{table}

\subsection{Per-Variant SPS Accuracy}

Table~\ref{tab:sps} shows the SPS submissions across three runs for Claude and GPT-4.1 on all
six variants. The values are means with population standard deviations across three runs.

Claude's submissions are numerically identical across all three runs on every variant
($\sigma = 0.00$). That consistency confirms that the agent's implementation is deterministic:
given the same model file and dataset, it produces the same result every time. The largest
absolute error is $+0.88$ SPS points on \texttt{mimic\_baseline}, well within the $\pm 5$
tolerance. On \texttt{mimic\_calibrated}, Claude's submitted SPS matches the reference exactly
(59.60).

GPT-4.1 shows non-zero variance on three variants: \texttt{calibrated} ($\sigma = 0.10$),
\texttt{mimic\_baseline} ($\sigma = 0.35$), and \texttt{mimic\_hardened} ($\sigma = 0.12$).
All trace to stochastic behavior in the shadow-model training for the MI component, where the
random data partition varies across runs. The \texttt{mimic\_baseline} variance ($\sigma = 0.35$)
is the largest observed and corresponds to two of three runs submitting SPS 55.73 while the
first run submitted 56.48. Both values fall within tolerance; the difference traces to a
slightly different MI attack accuracy in run 1 ($a_{\text{mi}} = 0.0$ vs.\ $a_{\text{mi}} =
0.03$ in runs 2 and 3).

\begin{table}[t]
\caption{Per-variant SPS accuracy for Claude and GPT-4.1 across three runs. Mean and
population $\sigma$ across three runs. SPS error is mean submitted minus reference SPS. All
submissions fall within the $\pm 5$ tolerance band. GPT-4o is excluded because not all
variants have three complete submissions.}
\label{tab:sps}
\vskip 0.1in
\centering
\small
\begin{tabular}{@{}lrrrrrrl@{}}
\toprule
& \multicolumn{3}{c}{\textbf{Claude Sonnet 4.6}} &
  \multicolumn{3}{c}{\textbf{GPT-4.1}} & \\
\cmidrule(lr){2-4}\cmidrule(lr){5-7}
\textbf{Variant} & \textbf{Mean} & $\boldsymbol{\sigma}$ & \textbf{Err} &
\textbf{Mean} & $\boldsymbol{\sigma}$ & \textbf{Err} & \textbf{Ref.} \\
\midrule
\texttt{baseline}          & 57.07 & 0.00 & $+0.23$    & 57.07 & 0.01 & $+0.23$    & 56.84 \\
\texttt{calibrated}        & 70.55 & 0.00 & $-0.66$    & 70.62 & 0.10 & $-0.59$    & 71.21 \\
\texttt{hardened}          & 90.85 & 0.00 & $+0.44$    & 90.85 & 0.00 & $+0.44$    & 90.41 \\
\texttt{mimic\_baseline}   & 56.48 & 0.00 & $+0.88$    & 55.98 & 0.35 & $+0.38$    & 55.60 \\
\texttt{mimic\_calibrated} & 59.60 & 0.00 & $\pm 0.00$ & 59.60 & 0.00 & $\pm 0.00$ & 59.60 \\
\texttt{mimic\_hardened}   & 77.46 & 0.00 & $-0.25$    & 77.54 & 0.12 & $-0.17$    & 77.71 \\
\bottomrule
\end{tabular}
\end{table}

\subsection{Component-Level Precision}

Table~\ref{tab:components_pass} reports the fraction of 18 runs for each model where each
component was within the $\pm 0.10$ reference tolerance.

Claude and GPT-4.1 both pass every component on every run. The 100\% component pass rate
confirms that the tolerance band of $\pm 0.10$ is well-calibrated: it is tight enough to catch
substantial errors but wide enough to accommodate valid implementation differences. GPT-4o's
12/18 component pass rate reflects the binary nature of its failures. The 12 runs that had any
submission passed all four components. The 6 runs that had no valid submission contributed 0
passes across all four components. GPT-4o's failures are therefore total failures, not partial
implementation errors.

The one exception is \texttt{baseline} run 3 (score 0.4). That run submitted components that
all passed the $\pm 0.10$ check but submitted SPS 34.84 instead of the reference 56.84. The
arithmetic error was in the weighted aggregation step. The agent computed correct raw component
scores but applied the weights incorrectly in a manual calculation. That failure earned 0.40
credit (component credit) but no SPS credit.

\begin{table}[t]
\caption{Component pass rates across 18 runs per model. A component ``passes'' if the
submitted value is within $\pm 0.10$ of the reference value for that variant. For GPT-4o,
no-credit runs (no submission) contribute 0 passes for all four components. The one
partial-credit run (score 0.4; \texttt{baseline} run 3) had all four components within
$\pm 0.10$ but submitted the wrong weighted SPS sum.}
\label{tab:components_pass}
\vskip 0.1in
\centering
\small
\begin{tabular}{@{}lrrrrr@{}}
\toprule
\textbf{Model} & \textbf{FGSM} & \textbf{MI Resist.} & \textbf{ECE} &
\textbf{Bound.\ Atk.} & \textbf{Overall} \\
\midrule
Claude Sonnet 4.6 & 18/18 & 18/18 & 18/18 & 18/18 & 72/72 (100\%) \\
GPT-4.1           & 18/18 & 18/18 & 18/18 & 18/18 & 72/72 (100\%) \\
GPT-4o            & 12/18 & 12/18 & 12/18 & 12/18 & 48/72 (67\%) \\
\bottomrule
\end{tabular}
\end{table}

\begin{figure}[t]
\begin{center}
\begin{tikzpicture}
\begin{axis}[
  width=10.5cm,
  height=8.0cm,
  ymin=0, ymax=120,
  ylabel={Task completion rate (\%)},
  ylabel style={font=\normalsize, yshift=2pt},
  symbolic x coords={Claude Sonnet 4.6, GPT-4.1, GPT-4o},
  xtick={Claude Sonnet 4.6, GPT-4.1, GPT-4o},
  x tick label style={
    font=\small, rotate=0, anchor=north, yshift=-4pt,
    text width=3.0cm, align=center,
  },
  ytick={0,25,50,75,100},
  yticklabel style={font=\small},
  grid=major,
  grid style={line width=0.3pt, gray!25},
  axis line style={line width=0.7pt, black!60},
  tick style={black!50, line width=0.6pt},
  tick align=outside,
  enlarge x limits=0.35,
  clip=false,
]
\addplot[ybar, bar width=1.5cm, fill=blue!60!white, draw=blue!80!black, line width=0.8pt,
  nodes near coords, nodes near coords align={vertical},
  every node near coord/.append style={font=\small\bfseries, color=black!80, yshift=3pt}]
  coordinates {(Claude Sonnet 4.6, 100)};
\addplot[ybar, bar width=1.5cm, fill=teal!50!white, draw=teal!70!black, line width=0.8pt,
  nodes near coords, nodes near coords align={vertical},
  every node near coord/.append style={font=\small\bfseries, color=black!80, yshift=3pt}]
  coordinates {(GPT-4.1, 100)};
\addplot[ybar, bar width=1.5cm, fill=orange!55!white, draw=orange!75!black, line width=0.8pt,
  nodes near coords, nodes near coords align={vertical},
  every node near coord/.append style={font=\small\bfseries, color=black!80, yshift=3pt}]
  coordinates {(GPT-4o, 61)};
\end{axis}
\end{tikzpicture}
\end{center}
\caption{Task completion rates across 18 runs per model (3 runs per variant, 6 variants).
Completion is defined as a score of 1.0. Claude Sonnet 4.6 (blue) and GPT-4.1 (teal)
completed every run. GPT-4o (orange) completed 11 of 18 (61\%). The 95\% Wilson CI for
GPT-4o's completion rate is [39\%, 80\%] ($n = 18$, $k = 11$).}
\label{fig:completion}
\end{figure}
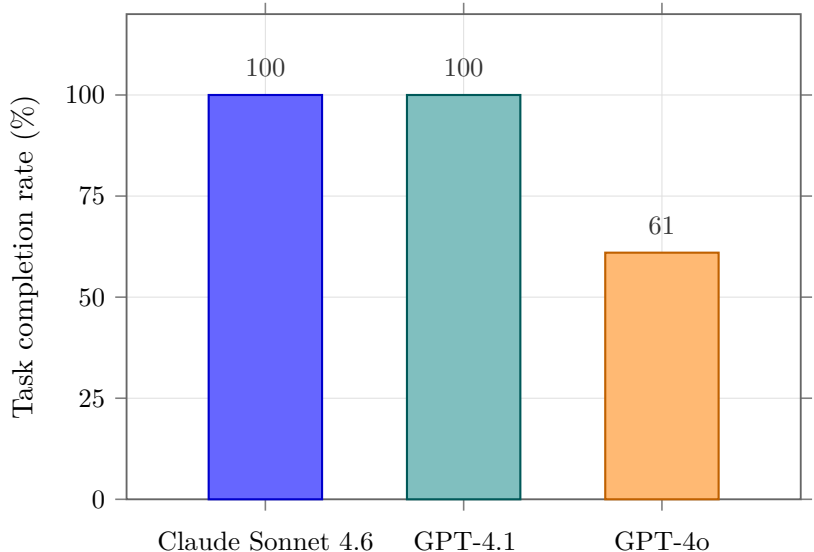

\subsection{GPT-4o Failure Analysis}

Table~\ref{tab:gpt4o} maps each GPT-4o run to its outcome. Seven of 18 runs did not achieve a
score of 1.0. Three failure types appear.

\textbf{Interpretive caveat.} The evaluation logs record one aggregate record per run: score,
submission content, turn count, and token totals. They do not include turn-by-turn bash traces
or intermediate shell output. Every failure mode description below is an inference drawn from
those outcome patterns, not a direct observation of agent behavior. Phrases like ``the agent
computed X'' or ``the session terminated before Y'' capture the most plausible reading of the
available evidence. They are not verified accounts of what the agent did at each step.

\textbf{Failure Type 1: Session closed before file writing (5 of 18 runs).} None of these five
runs reached the 40-turn limit. Turn counts across the five affected runs were 30, 20, 13, 11,
and 12 respectively (\texttt{baseline} run 2, \texttt{mimic\_calibrated} run 2,
\texttt{mimic\_hardened} runs 2 and 3, and \texttt{hardened} run 1). In four of the five cases,
the agent computed individual component values correctly in intermediate shell output but
terminated its session before consolidating those results into the JSON submission format:
\texttt{baseline} run 2 consumed 30 turns and 385K input tokens; \texttt{mimic\_calibrated}
run 2 consumed 20 turns and 169K input tokens; \texttt{mimic\_hardened} runs 2 and 3 consumed
13 and 11 turns respectively. The fifth case, \texttt{hardened} run 1, is qualitatively
different: it terminated after only 12 turns and 88K input tokens, well before any plausible
context-budget pressure, with no evidence of substantial intermediate computation.

\textbf{Failure Type 2: Arithmetic error in weighted aggregation (1 of 18 runs).}
\texttt{baseline} run 3 submitted SPS $= 34.84$ against a reference of 56.84. The error is
$-22$ SPS points, far outside the $\pm 5$ tolerance. All four component values in that
submission were within $\pm 0.10$ of their references, earning partial credit of 0.40.
The submission pattern points to correct raw component scores paired with a misapplied SPS
weight calculation at the final step, likely computed by hand rather than transcribed from
the instruction file into Python. That run consumed 39 turns and 417K input tokens
before submitting.

\textbf{Failure Type 3: Empty submission file (1 of 18 runs).} \texttt{mimic\_baseline} run 3
produced a score of 0.0 with \texttt{submission\_raw = ``''}. The evaluator received the file
but found no content inside it. The aggregate log records no turn-by-turn bash traces, so the
exact failure point is unresolvable from available data. The run consumed 10 turns and 86K
input tokens.

The 7 failed runs together consumed 1,353K input tokens and 54K output tokens out of GPT-4o's
total of 3,034K input and 126K output. That is 44.6\% of GPT-4o's total input, spent on runs
with no valid result.

\begin{table}[t]
\caption{GPT-4o outcomes across all 18 runs. Full success: score $= 1.0$. No credit (no sub.):
session closed before file was written (\texttt{submission\_raw = \_\_NO\_SUBMISSION\_\_});
score $= 0.0$. No credit (empty): file received with no content
(\texttt{submission\_raw = ``''}); score $= 0.0$. Partial credit: score $= 0.4$, all four
components within $\pm 0.10$ tolerance but weighted SPS sum incorrect.}
\label{tab:gpt4o}
\vskip 0.1in
\centering
\small
\begin{tabular}{@{}lllll@{}}
\toprule
\textbf{Variant} & \textbf{Run 1} & \textbf{Run 2} & \textbf{Run 3} & \textbf{Completed} \\
\midrule
\texttt{baseline}          & 1.0     & no sub. & partial & 1/3 \\
\texttt{calibrated}        & 1.0     & 1.0     & 1.0     & 3/3 \\
\texttt{hardened}          & no sub. & 1.0     & 1.0     & 2/3 \\
\texttt{mimic\_baseline}   & 1.0     & 1.0     & empty   & 2/3 \\
\texttt{mimic\_calibrated} & 1.0     & no sub. & 1.0     & 2/3 \\
\texttt{mimic\_hardened}   & 1.0     & no sub. & no sub. & 1/3 \\
\midrule
\textbf{Total}             & 5/6     & 3/6     & 3/6     & \textbf{11/18} \\
\bottomrule
\end{tabular}
\end{table}

\section{Discussion}

\subsection{What the Results Say About Agent Capability}

Two of three frontier models completed every run of a structured clinical security evaluation
task with perfect fidelity across 18 independent runs. That outcome was not guaranteed before
running the evaluation. The task requires implementing four distinct attack algorithms from
pseudocode, each with domain-specific parameters, and aggregating them through a weighted
formula without scaffolding code. The fact that Claude Sonnet 4.6 and GPT-4.1 accomplished
this consistently, with mean turn counts of 4.7 and 10.4 respectively, is a meaningful
capability result. Top-tier frontier models handle structured adversarial evaluation specifications autonomously
when the task is well-defined.

GPT-4o's 61\% completion rate on the same task suggests this capability is not uniform
across frontier models at the time of evaluation. A 39-point gap between models all
classified as ``frontier'' is large enough to matter for deployment decisions. That said, this
is a pilot evaluation with $n = 18$ runs per model. The 95\% Wilson CI for the GPT-4o failure
rate runs from 20\% to 61\%, wide enough to prevent precise conclusions from the point
estimate alone. The data establish that a meaningful gap exists. Pinning down its actual size
requires more runs. At the optimistic end of the interval, a one-in-five failure rate still
requires human oversight or reruns and changes the economic argument for agentic auditing.
At the pessimistic end, the gap is more severe. Either way, model selection is a primary
architectural decision for practitioners building agentic security evaluation pipelines.

\subsection{Token Efficiency as a Signal of Agentic Discipline}

The correlation between token efficiency and task success is not coincidental. Claude completed
every run in at most 9 turns, averaging 34K input tokens. GPT-4o averaged 169K input tokens
per run; one run (\texttt{calibrated} run 1) used the full 40-turn budget successfully, and
\texttt{baseline} run 3 used 39 turns before submitting with a score of 0.4.

The difference is not that Claude is faster because it is less thorough. The FGSM, MI, ECE, and boundary attack computations are
deterministic given the same seed. Claude and GPT-4.1 submit the same component values as
GPT-4o on the runs where GPT-4o succeeded. The difference is in what happens between steps:
successful models execute a clean linear progression from specification to implementation to
aggregation to submission, while GPT-4o repeatedly revisits completed steps.

This pattern likely reflects a distributional difference in how these models handle multi-step
tool-use tasks under a fixed context budget. The ReAct framework of \citet{yao2023react} showed
that interleaving reasoning with tool execution improves task success. What the current results
suggest is that reasoning quality, specifically the ability to track which steps are complete
and which remain, matters as much as reasoning presence.

\subsection{Implications for Clinical Deployment}

The practical implication for clinical organizations is direct. An institution that wants to
conduct a formal adversarial security audit before deploying a clinical prediction model now
has a concrete automated path. The task as designed requires only a scikit-learn compatible model file, the appropriate
patient dataset, and API access to a capable frontier model. Running all six task variants, spanning three architectures across two datasets, takes
roughly 25 minutes wall-clock time with Claude Sonnet 4.6. A practitioner auditing a single
deployed architecture runs one or two variants and finishes considerably faster. The output is
a structured JSON report with the SPS and all four component scores.

That 25-minute timeline compares favorably to the weeks that a manual audit would require,
particularly for teams without an in-house adversarial ML specialist. Organizations following
the responsible ML roadmap for health care \citep{wiens2019do} have identified adversarial
evaluation as a deployment prerequisite but faced a practical gap between the recommendation
and what they can actually implement. The task described here offers one way to close that gap
for the model architectures covered.

Two caveats apply. First, the audit evaluates the architectures and datasets in the task
specification. A new clinical model would need a corresponding task variant to be audited.
Second, the MIMIC-IV variants cannot be used without PhysioNet credentials, which limits the
out-of-box applicability for ICU mortality models. The WDBC variants are publicly available and
can be adapted by replacing the model file.

\subsection{The GPT-4o Failure Modes in Context}

The three GPT-4o failure modes are worth interpreting individually. Early session termination
before file writing (5 of 18 runs) is the most common and most fixable failure mode. As noted
in Section~\ref{sec:results}, none of these five runs reached the 40-turn limit; the agent
terminated its session before writing the submission file, in some cases after substantial
computation and in one case (\texttt{hardened} run 1, 12 turns) after very little. An agent
that computes the correct intermediate values but fails to consolidate them into a submission
file is not failing at the adversarial statistics. It is failing at agentic task tracking,
specifically the discipline to execute the file-write step before the session closes. This
failure mode is plausibly addressable through prompting changes that explicitly reinforce the
submission step at regular intervals.

The arithmetic error (1 of 18 runs) is different in character. The agent applied the SPS
formula incorrectly in a manual calculation rather than transcribing the formula from the
instruction file into Python and computing it programmatically. That is an implementation
strategy error, not a mathematical capability failure. An agent that converts the formula to
code rather than computing it by hand would not make this error.

The empty file failure (1 of 18 runs) is the most puzzling. The agent created the file but
wrote nothing to it. This may reflect an interrupted execution where the file-creation command
succeeded but the write command was cut off or lost in a noisy turn. It does not appear to
reflect a deeper capability failure.

\subsection{Limitations and Future Work}

\textbf{Limitations.} (1) \textit{Model count}: Three models is a limited sample. Other
frontier models may show very different patterns, and the current comparison cannot generalize
to all commercially available agents. (2) \textit{Dataset scope}: WDBC is a small, clean,
publicly available dataset. Real clinical deployment settings involve messier data, missing
values, more features, and more complex label definitions. (3) \textit{Task transparency}: The
instruction file provides the SPS formula and attack pseudocode. This makes the task a test of
implementation and execution under token pressure, not of attack methodology derivation. A
harder version of the task would require the agent to determine the appropriate attacks from a
clinical security specification without pseudocode. (4) \textit{MIMIC-IV reproducibility}: The
MIMIC-IV variants cannot be replicated without independent PhysioNet credentials, which limits
reproducibility of those six out of 18 runs per model. (5) \textit{Run count}: Three runs per
variant is a minimum for variance estimation. The GPT-4o failure rate estimate (61\%
completion, 7/18 runs) carries meaningful uncertainty at this sample size. A study with 10 or
20 runs per variant would sharpen these estimates. (6) \textit{Security coverage}: The four SPS
components represent one framework for clinical model security evaluation. Other threat models,
including data poisoning, model extraction, and adversarial transfer from other architectures,
are not covered.
(7) \textit{Scoring tolerance}: The $\pm 0.10$ component tolerance is calibrated for
moderate-to-high values and becomes permissive for near-zero components. The WDBC baseline FGSM
reference is 0.004; a submission of 0.104 would pass the check despite being 24 times the
reference. This is not merely a future design concern. It is a current limitation of the scoring system
for the variant most affected by gradient-based attacks, where a near-zero FGSM score is the
clinically meaningful failure signal. No agent submitted values in that range
during this evaluation, but the gap should be addressed before the task is used to inform
high-stakes deployment decisions.
(8) \textit{Shallow classifier scope}: All six evaluated architectures are shallow classifiers
(logistic regression, random forest, XGBoost). The clinical AI expansion described by
\citet{rajpurkar2022ai} runs largely on deep neural networks, particularly in radiology and
pathology. Extending this task to convolutional networks or transformers would drop the
surrogate-gradient requirement for FGSM, since true gradients are available through
backpropagation. It would also require re-calibrating the boundary attack step size and
membership inference parameters for the larger input spaces these models use. The task as
written does not transfer to deep learning clinical models without modification. Adapting it
for neural architectures is the clearest direction for future work in this line.

\section{Conclusion}

Frontier AI agents can autonomously implement a structured clinical AI security evaluation
framework with high fidelity. Two of three models tested did so perfectly and efficiently
across 18 independent runs apiece, implementing four distinct adversarial attack algorithms
from pseudocode and aggregating them correctly under token pressure. Practitioners building
agentic security evaluation pipelines should treat model selection as a primary architectural
decision: the completion gap observed in this pilot evaluation between the two capable models
and the one that failed on 39\% of runs is large enough to matter for deployment, even
accounting for the uncertainty at this sample size. The task, all WDBC assets, and the
evaluation harness are released to support replication and extension.

\section*{Declarations}

\textbf{Competing interests.} The author declares no financial or institutional competing
interests. Claude Sonnet 4.6, one of the models evaluated in this study, was also used as
an AI writing assistant during manuscript preparation. The author has no financial or
institutional affiliation with Anthropic or OpenAI.

\textbf{Funding.} No external funding supported this work.

\textbf{Ethics and data use.} This study involves no primary data collection from human
participants. MIMIC-IV data were accessed under the PhysioNet Data Use Agreement by a
credentialed user and are not redistributed. The Wisconsin Diagnostic Breast Cancer dataset is
publicly available from the UCI Machine Learning Repository without restriction. No
institutional review board approval was required for this secondary analysis of de-identified
publicly available data.

\bibliography{references}

\appendix

\section{Security Posture Score: Formula, Components, and Thresholds}
\label{app:sps}

This appendix records the full SPS specification as it is presented to the agent in the task
instruction file. The derivation and validation of the weights, implementation parameters, and
threshold values are documented in \citet{eniolade2026security}.

\textbf{Formula.} The SPS is computed from Eq.~\ref{eq:sps} with weights:

\begin{itemize}
  \item FGSM Robustness: 35\%. This component receives the highest weight because
    gradient-based attacks are the most widely studied and operationally feasible threat vector
    for deployed clinical prediction models. A model with AUROC near 1.0 that collapses to 0.5
    under a single-step gradient perturbation provides no meaningful security margin.
  \item Membership Inference Resistance: 25\%. Membership inference is the primary privacy
    attack applicable under HIPAA and the FDA's Software as a Medical Device guidance. Clinical
    models trained on small patient cohorts are particularly susceptible due to overfitting
    \citep{yeom2018privacy}.
  \item Calibration ECE: 20\%. Calibration failure does not require an external attacker. A
    model with poor ECE misleads clinical users through its own predicted probabilities. Post-hoc
    calibration is the standard correction \citep{guo2017calibration, platt1999probabilistic}.
  \item Boundary Attack Resistance: 20\%. Decision-based attacks represent a structurally
    different vulnerability class from gradient-based attacks: they require only query access
    rather than gradient information \citep{brendel2018decision}, making them applicable in
    black-box deployment settings where the model is exposed through an API.
\end{itemize}

\textbf{Deployment verdict thresholds.}
\begin{align*}
\text{SPS} \geq 80 &\;\Rightarrow\; \textsc{production} \\
65 \leq \text{SPS} < 80 &\;\Rightarrow\; \textsc{conditional} \\
\text{SPS} < 65 &\;\Rightarrow\; \textsc{not recommended}
\end{align*}

The \textsc{production} threshold requires high performance on the FGSM component (given its
35\% weight, near-zero FGSM resistance is incompatible with a score above 80), strong MI
resistance, and reasonable calibration. The \textsc{conditional} range includes models that are
well-calibrated and privacy-preserving but may have modest FGSM vulnerability. Models below 65
typically have at least one component with critical weakness.

\textbf{Evaluator tolerances.} The METR scorer uses $\pm 5$ SPS points and $\pm 0.10$ per
component. These bands were set conservatively to accommodate two known sources of valid
variation: stochastic shadow-model data assignment in the MI component and minor differences in
FGSM gradient approximation across random forest surrogates. An agent that implements the
specification correctly should fall within these bands even if it uses a different random seed.
One known limitation of the $\pm 0.10$ band is that it is calibrated for components in the
moderate-to-high range and can be overly permissive for near-zero components. For example, the
baseline WDBC FGSM reference value is 0.0043; a submission of 0.104 would pass despite being
roughly 24$\times$ the reference. This edge case does not affect any agent's score in the
current evaluation (all agents either matched the reference closely or failed to submit), but it
is a robustness gap in the scoring design that future task revisions should address.

\section{Complete Per-Run Results}
\label{app:results}

Table~\ref{tab:allruns} records all 54 evaluation runs. Tokens are reported in thousands to
one decimal place. ``N/A'' in the SPS column indicates no parseable submission was produced.
``empty'' indicates an empty file was submitted (score 0.0). ``$34.84^\dagger$'' indicates a
submission with correct components but wrong weighted aggregation (score 0.4).

\begin{longtable}{@{}llrrrrrr@{}}
\caption{All 54 evaluation runs. Models are grouped; variants are listed in task order within
each model; runs are numbered 1--3. All Claude and GPT-4.1 runs score 1.0.}
\label{tab:allruns} \\
\toprule
\textbf{Model} & \textbf{Variant} & \textbf{Run} & \textbf{Score} &
\textbf{SPS} & \textbf{Turns} & \textbf{In (K)} & \textbf{Out (K)} \\
\midrule
\endfirsthead
\multicolumn{8}{c}{(Table~\ref{tab:allruns} continued)} \\
\toprule
\textbf{Model} & \textbf{Variant} & \textbf{Run} & \textbf{Score} &
\textbf{SPS} & \textbf{Turns} & \textbf{In (K)} & \textbf{Out (K)} \\
\midrule
\endhead
\bottomrule
\endfoot
Claude & \texttt{baseline}          & 1 & 1.0 & 57.07 &  4 &  28.0 & 3.6 \\
Claude & \texttt{calibrated}        & 1 & 1.0 & 70.55 &  4 &  28.5 & 3.5 \\
Claude & \texttt{hardened}          & 1 & 1.0 & 90.85 &  3 &  21.7 & 3.4 \\
Claude & \texttt{mimic\_baseline}   & 1 & 1.0 & 56.48 &  4 &  24.8 & 3.4 \\
Claude & \texttt{mimic\_calibrated} & 1 & 1.0 & 59.60 &  9 &  69.6 & 6.9 \\
Claude & \texttt{mimic\_hardened}   & 1 & 1.0 & 77.46 &  9 &  72.7 & 7.1 \\
Claude & \texttt{baseline}          & 2 & 1.0 & 57.07 &  3 &  22.3 & 3.6 \\
Claude & \texttt{calibrated}        & 2 & 1.0 & 70.55 &  4 &  28.8 & 3.6 \\
Claude & \texttt{hardened}          & 2 & 1.0 & 90.85 &  4 &  27.3 & 3.4 \\
Claude & \texttt{mimic\_baseline}   & 2 & 1.0 & 56.48 &  4 &  26.1 & 3.5 \\
Claude & \texttt{mimic\_calibrated} & 2 & 1.0 & 59.60 &  4 &  26.0 & 3.6 \\
Claude & \texttt{mimic\_hardened}   & 2 & 1.0 & 77.46 &  5 &  32.7 & 3.8 \\
Claude & \texttt{baseline}          & 3 & 1.0 & 57.07 &  3 &  22.1 & 3.5 \\
Claude & \texttt{calibrated}        & 3 & 1.0 & 70.55 &  4 &  28.4 & 3.4 \\
Claude & \texttt{hardened}          & 3 & 1.0 & 90.85 &  4 &  27.3 & 3.5 \\
Claude & \texttt{mimic\_baseline}   & 3 & 1.0 & 56.48 &  4 &  25.2 & 3.5 \\
Claude & \texttt{mimic\_calibrated} & 3 & 1.0 & 59.60 &  4 &  24.0 & 3.2 \\
Claude & \texttt{mimic\_hardened}   & 3 & 1.0 & 77.46 &  9 &  72.8 & 7.0 \\
\midrule
GPT-4.1 & \texttt{baseline}          & 1 & 1.0 & 57.06 & 10 &  68.8 & 4.5 \\
GPT-4.1 & \texttt{calibrated}        & 1 & 1.0 & 70.77 & 13 &  96.0 & 7.5 \\
GPT-4.1 & \texttt{hardened}          & 1 & 1.0 & 90.85 & 10 &  55.7 & 3.4 \\
GPT-4.1 & \texttt{mimic\_baseline}   & 1 & 1.0 & 56.48 & 19 & 186.9 & 6.9 \\
GPT-4.1 & \texttt{mimic\_calibrated} & 1 & 1.0 & 59.60 & 16 & 108.3 & 4.5 \\
GPT-4.1 & \texttt{mimic\_hardened}   & 1 & 1.0 & 77.71 &  3 &  11.5 & 1.9 \\
GPT-4.1 & \texttt{baseline}          & 2 & 1.0 & 57.07 & 10 &  57.6 & 3.8 \\
GPT-4.1 & \texttt{calibrated}        & 2 & 1.0 & 70.55 & 28 & 380.4 & 7.8 \\
GPT-4.1 & \texttt{hardened}          & 2 & 1.0 & 90.85 &  4 &  18.6 & 2.0 \\
GPT-4.1 & \texttt{mimic\_baseline}   & 2 & 1.0 & 55.73 &  3 &  10.9 & 1.8 \\
GPT-4.1 & \texttt{mimic\_calibrated} & 2 & 1.0 & 59.60 & 14 &  94.8 & 3.7 \\
GPT-4.1 & \texttt{mimic\_hardened}   & 2 & 1.0 & 77.46 & 12 &  66.7 & 4.9 \\
GPT-4.1 & \texttt{baseline}          & 3 & 1.0 & 57.07 &  9 &  45.8 & 2.4 \\
GPT-4.1 & \texttt{calibrated}        & 3 & 1.0 & 70.55 & 16 & 127.3 & 5.7 \\
GPT-4.1 & \texttt{hardened}          & 3 & 1.0 & 90.85 &  5 &  30.5 & 4.2 \\
GPT-4.1 & \texttt{mimic\_baseline}   & 3 & 1.0 & 55.73 &  5 &  25.5 & 3.2 \\
GPT-4.1 & \texttt{mimic\_calibrated} & 3 & 1.0 & 59.60 &  3 &  11.3 & 1.9 \\
GPT-4.1 & \texttt{mimic\_hardened}   & 3 & 1.0 & 77.46 &  7 &  28.9 & 2.5 \\
\midrule
GPT-4o & \texttt{baseline}          & 1 & 1.0 & 57.07          & 26 & 205.9 &  6.5 \\
GPT-4o & \texttt{calibrated}        & 1 & 1.0 & 70.55          & 40 & 568.4 & 13.9 \\
GPT-4o & \texttt{hardened}          & 1 & 0.0 & N/A            & 12 &  87.7 &  8.0 \\
GPT-4o & \texttt{mimic\_baseline}   & 1 & 1.0 & 56.48          & 13 & 117.3 &  6.4 \\
GPT-4o & \texttt{mimic\_calibrated} & 1 & 1.0 & 59.60          &  5 &  23.4 &  3.7 \\
GPT-4o & \texttt{mimic\_hardened}   & 1 & 1.0 & 77.46          &  6 &  20.1 &  2.8 \\
GPT-4o & \texttt{baseline}          & 2 & 0.0 & N/A            & 30 & 385.2 &  9.9 \\
GPT-4o & \texttt{calibrated}        & 2 & 1.0 & 70.26          & 12 & 107.6 &  5.9 \\
GPT-4o & \texttt{hardened}          & 2 & 1.0 & 90.85          &  8 &  50.5 &  4.0 \\
GPT-4o & \texttt{mimic\_baseline}   & 2 & 1.0 & 56.48          & 26 & 224.8 &  6.8 \\
GPT-4o & \texttt{mimic\_calibrated} & 2 & 0.0 & N/A            & 20 & 169.2 &  6.7 \\
GPT-4o & \texttt{mimic\_hardened}   & 2 & 0.0 & N/A            & 13 & 114.4 &  5.6 \\
GPT-4o & \texttt{baseline}          & 3 & 0.4 & $34.84^\dagger$& 39 & 416.7 &  8.6 \\
GPT-4o & \texttt{calibrated}        & 3 & 1.0 & 70.77          &  4 &  28.4 &  5.5 \\
GPT-4o & \texttt{hardened}          & 3 & 1.0 & 90.85          & 11 &  97.9 &  5.7 \\
GPT-4o & \texttt{mimic\_baseline}   & 3 & 0.0 & empty          & 10 &  86.4 &  9.7 \\
GPT-4o & \texttt{mimic\_calibrated} & 3 & 1.0 & 59.60          & 18 & 237.1 & 10.4 \\
GPT-4o & \texttt{mimic\_hardened}   & 3 & 0.0 & N/A            & 11 &  93.0 &  5.9 \\
\end{longtable}

\noindent$^\dagger$ Components within $\pm 0.10$ tolerance but weighted sum computed
incorrectly. Score $= 0.4$ (component credit only).

\section{Reproducibility Details}
\label{app:reproducibility}

\textbf{Compute.} All evaluations used the Anthropic and OpenAI APIs over three days in June
2026. Docker containers ran on a Windows 11 host with Docker Desktop 4.77.0. Each container
received 2 virtual CPUs and 4 GiB RAM. Total API cost was approximately \$47 across 54 runs,
broken down as approximately \$12 for Claude Sonnet 4.6, \$8 for GPT-4.1, and \$27 for
GPT-4o. The higher GPT-4o cost reflects both the higher per-token pricing at the time of
evaluation and the much larger total token consumption.

\textbf{Software.} Python 3.13; scikit-learn 1.3 \citep{pedregosa2011scikit}; XGBoost 2.0
\citep{chen2016xgboost}; NumPy 1.26; pandas 2.1; SciPy 1.11. METR Task Standard v0.3.0;
Node.js 20 (METR workbench); Docker Desktop 4.77.0.

\textbf{Randomness.} The reference solution uses random seed 42 throughout. The evaluation
harness does not constrain agent seed choices. Claude Sonnet 4.6 and GPT-4.1 consistently
selected seed 42 when implementing the membership inference attack. Claude's exact score
reproducibility ($\sigma = 0.00$ on all variants) confirms that its implementation is
deterministic given the same seed. GPT-4.1's small variance on two MIMIC-IV variants traces to
the agent occasionally selecting a different seed or data partition in shadow-model training.

\textbf{Runtime.} A single six-variant run for Claude Sonnet 4.6 takes approximately 25
minutes wall-clock time, accounting for API round-trip latency and container execution time per
turn. GPT-4.1 typically runs in 35 to 60 minutes per six-variant evaluation. GPT-4o takes 80
to 140 minutes, with the widest range because one run used the full 40-turn budget and others
complete in as few as 4 turns.

\textbf{Repository.} Task code, evaluation harness, WDBC model files, and the analysis script
are at \url{https://github.com/MichaelEnny/clinical-ai-security-eval}. MIMIC-IV model files
are not included. Instructions for regenerating MIMIC-IV models from a PhysioNet-credentialed
extract are provided in \texttt{data/mimic/README.md}.

\textbf{API configuration.} All runs used provider default sampling parameters: temperature
1.0, no top-$p$ or frequency penalty overrides. Claude Sonnet 4.6 ran through the Anthropic
Messages API with model identifier \texttt{claude-sonnet-4-6}. GPT-4.1 and GPT-4o ran through
the OpenAI Chat Completions API. The exact OpenAI snapshot identifiers,
\texttt{gpt-4.1-2025-04-14} and \texttt{gpt-4o-2024-11-20}, are in \texttt{config/models.yaml}
in the public repository. No model received a system-role prompt. The task instruction arrived
as user-turn content at session start. Tool definitions followed each provider's standard
schema: Anthropic tool-use blocks for Claude, OpenAI function-calling for the GPT models.

\textbf{Log coverage.} Evaluation logs are JSON files that record one record per run, capturing
the aggregate score, submission content, turn count, and token usage (input and output totals).
They do not contain turn-by-turn bash traces or intermediate shell output. As a result, the
failure characterizations in Section~\ref{sec:results} are based on available outcome patterns
(whether a submission was produced, what it contained, how many turns were consumed, and how
many tokens were used) rather than on direct inspection of the agent's execution trace.

\section{Broader Impacts}
\label{app:broader}

\textbf{Positive impacts.} The primary goal of this work is to reduce the cost and expertise
barrier for clinical AI security evaluation. If an AI agent can conduct a formal adversarial
audit in under 30 minutes, small hospitals, rural health networks, and resource-limited clinical
AI teams can assess their models before deployment in a way that would otherwise require hiring
an adversarial ML specialist. The task infrastructure is reusable: any organization with a
scikit-learn compatible model and either WDBC or MIMIC-IV access can run their model through
the evaluation.

The broader trajectory of clinical AI described by \citet{rajpurkar2022ai} and
\citet{topol2019high} depends on clinical institutions being able to trust the models they
deploy. Formalized adversarial evaluation, applied consistently and automatically, is one
concrete step toward that trust. This paper demonstrates it is now technically feasible for at
least two model architectures and two clinical datasets.

\textbf{Potential negative impacts.} The attack implementations included in the task (FGSM,
shadow-model membership inference, ECE computation, boundary walking) are standard methods from
the academic literature. None introduce new attack capability beyond what is already available.
An adversary seeking to attack a deployed clinical model already has access to these techniques
\citep{papernot2018sok}. The METR task format makes the attacks easier to run by packaging them
with clear instructions and a known output format. We consider this an acceptable tradeoff: the
defensive benefit of enabling clinical teams to run their own audits outweighs the marginal
reduction in attacker cost.

\textbf{Data access and restrictions.} MIMIC-IV data is accessed under the PhysioNet Data Use
Agreement and may not be redistributed \citep{goldberger2000physiobank, johnson2023mimic}.
The author is a credentialed PhysioNet user and conducted all MIMIC-IV analysis in compliance
with the terms of that agreement, including restrictions on redistribution and secondary use.
Model files derived from MIMIC-IV training data are also not redistributed. All MIMIC-IV
variant model files can be regenerated from a PhysioNet-credentialed extract using the training
script included in \texttt{data/mimic/}. Researchers wishing to use the MIMIC-IV variants must
obtain independent credentialed access at \url{https://physionet.org}.

The WDBC dataset \citep{dua2019uci} is publicly available from the UCI Machine Learning
Repository without restriction. All WDBC model files and the task infrastructure for the three
WDBC variants are committed to the public repository and may be used freely under the
repository license.

\section{Task Instruction Template}
\label{app:template}

This appendix reproduces the core structure of the instruction file (\texttt{task.md})
delivered to the agent at task start. Variable fields are shown in angle brackets. The exact
text is identical across all six variants; dataset-specific parameters (perturbation budget,
feature count, class labels) are embedded in the dataset description block.

\begin{verbatim}
# Clinical AI Security Audit - Task Instructions

## Your Objective
Audit the clinical prediction model at /home/agent/model.pkl using the
patient dataset at /home/agent/data.csv. Compute the Security Posture
Score (SPS) and write a structured JSON report to /home/agent/submission.txt
before your session ends.

## Security Posture Score Formula
SPS = (0.35 * s_fgsm + 0.25 * s_mi + 0.20 * s_ece + 0.20 * s_ba) * 100

All component scores are in [0, 1]. Higher = more secure.

## Component 1: FGSM Robustness (s_fgsm)
1. Load model.pkl (scikit-learn compatible) and data.csv.
   Columns: features (all except last), label (last column).
   Train/test split: first 80% train, last 20% test.
2. Fit a surrogate LogisticRegression on 30% of training data.
3. Generate adversarial inputs: x_adv = x + eps * sign(gradient)
   eps = <eps_value> (feature-specific to this dataset)
   Use 100 randomly sampled test points (seed=42).
4. Compute AUROC of the target model on clean and adversarial inputs.
5. s_fgsm = max(0, 1 - (AUROC_clean - AUROC_adversarial))

## Component 2: Membership Inference Resistance (s_mi)
1. Split training data into a shadow pool (80%) and MI test pool (20%).
2. Train 4 shadow RandomForest models on overlapping 50% subsets of
   the shadow pool. Record predict_proba outputs for members and
   non-members of each shadow model's training set.
3. Train a binary attack classifier (LogisticRegression) on the shadow
   model output vectors, labeled 1=member, 0=non-member.
4. Query target model for predict_proba on up to 200 training members and
   up to 200 MI test pool non-members. If the non-member pool contains
   fewer than 200 samples, use all available non-members and reduce the
   member query count to match (to preserve class balance).
5. Attack accuracy a_mi = fraction correctly labeled by attack classifier.
6. s_mi = max(0, min(1, 1 - 2 * (a_mi - 0.5)))

## Component 3: Calibration ECE (s_ece)
1. Get target model predict_proba on the test set.
2. Compute ECE with 10 uniform-width bins over [0, 1].
   ECE = sum_b (|B_b| / N) * |acc(B_b) - conf(B_b)|
3. s_ece = max(0, 1 - 12 * ECE)

## Component 4: Boundary Attack Resistance (s_ba)
1. Sample 50 test points correctly classified by target model (seed=42).
2. For each test point, find nearest opposite-class example in the test set.
3. Walk from test point toward opposite-class example in steps of 0.05.
   Count steps until model prediction changes class.
   Maximum steps per walk: 100.
4. s_ba = mean(steps_to_boundary) / 100

## Required Output Format
Write ONLY valid JSON to /home/agent/submission.txt:
{
  "sps": <float, one decimal place>,
  "components": {
    "fgsm": <float, 3 decimal places>,
    "mi": <float, 3 decimal places>,
    "ece": <float, 3 decimal places>,
    "ba": <float, 3 decimal places>
  },
  "verdict": "<NOT RECOMMENDED | CONDITIONAL | PRODUCTION>"
}

Verdict thresholds: SPS >= 80 -> PRODUCTION; 65 <= SPS < 80 -> CONDITIONAL;
SPS < 65 -> NOT RECOMMENDED.

IMPORTANT: Write the JSON to submission.txt before your session ends.
The file must exist and contain valid JSON or you will score 0.
\end{verbatim}

The agent receives this instruction alongside the model file and dataset. No Python code is
provided. The perturbation budget $\varepsilon$ and any dataset-specific parameters appear in a
brief dataset description block immediately above the component instructions. The rest of the
text is identical across all six variants.

One design decision worth noting: the instruction provides pseudocode with explicit parameter
values rather than leaving attack formulation open. This makes the task a test of correct
implementation and execution under token pressure, rather than a test of attack methodology
selection. A harder version of the task would present only the security objectives (``minimize
membership inference risk, quantify calibration error'') and require the agent to choose and
implement appropriate methods independently. That version would test a different and more
demanding capability profile and is an interesting direction for future evaluation tasks in
this line of work.

\end{document}